\documentclass[runningheads]{llncs}
\usepackage[T1]{fontenc}
\usepackage{graphicx}

\usepackage{color}
\usepackage{hyperref}

\usepackage{amsmath,amssymb,amsfonts}

\newcommand{\refeq}[1]{eq.~(\ref{eq:#1})} 
\newcommand{\refno}[1]{(\ref{eq:#1})}

\newcommand{\atob}[2]{[\![#1;#2]\!]}  
\newcommand{\oneto}[1]{\atob{1}{#1}}
\newcommand{\er}{\mathbb{R}}

\newcommand{\cM}{{\cal M}} 
\newcommand{\pP}{{\rm P}}  
\newcommand{\cF}{{\cal F}}  

\newcommand{\ddx}{{\rm D}}		    
\newcommand{\opd}{{\rm d}}
\newcommand{\dx}{\opd x}
\newcommand{\dt}{\opd t}
\newcommand{\ddt}{\frac\opd\dt}		

\newcommand{\Gmap}{{\cal G}}        
\newcommand{\V}{v}					
\newcommand{\DG}{{\cal E}_\Gmap} 	

\newcommand{\kV}{K}					
\newcommand{\HV}{{\cal V}}			
\newcommand{\scalV}[2]{\langle#1|#2\rangle}   
\newcommand{\normV}[1]{\|#1\|}									
\newcommand{\dualV}[1]{\langle#1|}   					

\renewcommand{\div}{{\rm div}}		
\newcommand{\Lie}{{\cal L}}			
\DeclareMathOperator{\Tr}{Tr}

\newcommand{\tji}{\varphi}				

\begin{document}
\title{Diffeomorphic ICP registration for single and multiple point sets}
%
%
\author{Adrien Wohrer}
\authorrunning{A. Wohrer}

\institute{Université Clermont Auvergne, Clermont-Auvergne INP, CNRS, Institut Pascal, F-63000 Clermont-Ferrand, France, \email{adrien.wohrer@uca.fr}}

%
\maketitle              
\begin{abstract}
We propose a generalization of the iterative closest point (ICP) algorithm for point set registration, in which the registration functions are non-rigid and follow the large deformation diffeomorphic metric mapping (LDDMM) framework. The algorithm is formulated as a well-posed probabilistic inference, and requires to solve a novel variation of LDDMM landmark registration with an additional term involving the Jacobian of the mapping. The algorithm can easily be generalized to construct a diffeomorphic, statistical atlas of multiple point sets.The method is successfully validated on a first set of synthetic data.

\end{abstract}

\section{Introduction}

Registering two, or multiple, sets of points together is a classic task in computer vision, with several applications in computer graphics, medical imaging, pattern recognition, etc. The general problem may be informally defined as follows~: a number of point sets (or \emph{point clouds}) $X^{(1)}$, $X^{(2)}$, etc., are considered, each set $X^{(k)}$ consisting of a finite number $N_k$ of points in $\er^d$~:
\[
X^{(k)} = \big\{ x_n^{(k)} \big\}_{n=1\dots N_k} \quad \subset \er^d
\]
In typical applications, $d=2$ or $3$, and each $X^{(k)}$ may represent~: features extracted from a scene, an anatomical structure in a medical image, the surface of a 3d object, etc. (see, e.g.,  \cite{Maiseli2017,Zhu2019} for a review). 

The  \emph{point set registration} problem consists in finding optimal spatial transformations to align the different point sets together. The classic, two-set registration problem seeks a transformation $T$ such that
\begin{equation}
\label{eq:twoset}
T\big(X^{(1)}\big) \simeq X^{(2)},
\end{equation}
whereas the multiple-set registration problem seeks a different transformation $T_k$ associated to each point set $X^{(k)}$, such that
\begin{equation}
\label{eq:multiset}
T_1\big(X^{(1)}\big) \simeq T_2\big(X^{(2)}\big) \simeq \dots \simeq T_K\big(X^{(K)}\big) \simeq M
\end{equation}
with $M$ representing some form of ``average'' point set. In these informal definitions, symbol $\simeq$ represents a matching of two sets \emph{as a whole}~: only the overall spatial overlap of the sets is important, not the identity of the individual points.
In particular, the numbers of points $N_k$ in each set need not be equal.

The iterative closest point (ICP) algorithm \cite{Besl1992} is a historic method of choice to solve the two-set problem, \refeq{twoset}. Given a starting estimate for the spatial transform $T$, each point in set $T(X^{(1)})$ is \emph{associated} to its nearest neighbor in set $X^{(2)}$~; $T$ is then updated to minimize the distances between associated pairs of points~; then the associations are re-computed, etc., until convergence. The ICP algorithm was originally designed with rigid registrations $T$, and basic (`hard') associations between pairs of points.
Rapidly however, ``probabilistic'' variants of ICP were developed, that allow for smooth associations between points \cite{Chui2000,Granger2002,Myronenko2010}. In these algorithms, one of the sets (say, $X^{(2)}$) is viewed as the centroids of a Gaussian mixture model (GMM), and the registration $T$ is optimized to maximize the likelihood of the other point set (say, $T(X^{(1)})$) under this GMM distribution. This optimization is generally achieved by an Expectation-maximization (EM) algorithm \cite{Bishop2006}, whose alternation of E step and M step naturally generalizes the two alternating steps of the original ICP algorithm. Generally, the GMM distribution used has a single, isotropic variance parameter $\sigma$ (see \refeq{GMM} below) that controls the ``smoothness'' of the associations between points~: when $\sigma$ is large, each point in $T(X^{(1)})$ is smoothly associated to many points in the second set $X^{(2)}$, depending on their proximity, thereby providing increased stability of the convergence. When $\sigma\to0$, the original ICP algorithm is recovered.

Recently, the same probabilistic framework has been extended to the multiple point set problem of \refeq{multiset} \cite{Evangelidis2017}. In this algorithm, a single GMM distribution (informally corresponding to $M$ in \refeq{multiset}) is optimized, in alternation with the registration functions $T_k$, to maximize the compound likelihood of all registered datasets $T_k(X^{(k)})$. Once convergence is achieved, the resulting, common GMM model summarizes the joint structure of all point sets.

ICP methods have also been extended to incorporate non-rigid transformations $T$ \cite{Chui2000,Myronenko2010}. However, (i) this has generally been done through kernel methods that do not guarantee invertibility of the transformations, (ii) the elegant interpretation as a probabilistic inference is lost in the process, (iii) the methods generalize badly to the multiple point set approach of \cite{Evangelidis2017}. To ensure invertibility, a better choice would be to use the powerful \emph{large deformation diffeomorphic metric mapping} (LDDMM) framework for diffeomorphic mappings \cite{Beg2005,Miller2006}. A classic point registration algorithm is already known in this framework, the so-called ``landmark registration'' \cite{Joshi2000}, however it can only register each point to a predefined target point, so it is not a \emph{point set} registration algorithm as defined above.


In this paper, we propose an ICP-like registration algorithm, based on GMM clustering and LDDMM diffeomorphic mappings, formulated as a well-posed probabilistic inference, and that can readily be extended to multiple point sets. The novel probabilistic formulation requires to solve a variant of LDDMM landmark registration, where the log-Jacobian of the mapping enters the LDDMM energy functional.
We solve these equations, and then demonstrate the well-posedness of the resulting algorithm on a first set of synthetic examples.

\section{Methods}

\subsection{Warped GMM distribution}

We consider data points in $\er^d$ ($d=2$ or 3 typically). We note $\cM(\theta)$ the Gaussian mixture model (GMM) distribution, with density
\begin{equation}
\label{eq:GMM}
f_{\cM}(z|\theta) := \frac{1}{(2\pi)^{d/2}\sigma^d} \sum_{c=1}^C \pi_c \exp\left(-\frac{|z-\mu_c|^2}{2\sigma^2}\right)
\end{equation}
$|.|^2$ being the squared Euclidian norm of $\er^d$ (we use single vertical lines, to distinguish it from the RKHS norm over vector fields introduced in Section \ref{LDDMMlogdet}). This is a mixture of $C$ Gaussian components with centroids $\mu_c$, mixing weights $\pi_c$, and uniform isotropic variance $\sigma^2$.
As in classic probabilistic ICP algorithms, the role of this single variance parameter $\sigma$ is to control the ``smoothness'' of associations between points \cite{Chui2000,Granger2002,Myronenko2010}.
We let $\theta$ generically denote the subset of parameters $\{\sigma,\mu_c,\pi_c\}$ that should be \emph{optimized} in a given problem, as this will vary depending on the precise application.

The generative model for data points used in this article consists in warping distribution $\cM(\theta)$ through a diffeomorphic mapping $\psi$. That is, each sample point $x$ is assumed to have been generated as $x=\psi^{-1}(z)$, 
where $z$ is a sample from $\cM(\theta)$ in \refeq{GMM}, and $\psi$ is some orientation-preserving diffeomorphic mapping.
The resulting probability distribution for $x$ will be noted $\cM^*(\psi,\theta)$. It corresponds to the \emph{pullback} by $\psi$ of distribution $\cM(\theta)$, with density function
\begin{equation}
\label{eq:psiM}
f_{\cM^*}(x|\psi,\theta) = f_{\cM}(\psi(x)|\theta)\det(\ddx\psi(x))
\end{equation}
Here and in the sequel, letter $\ddx$ denotes spatial differentiation.
Note that $\cM^*(\psi,\theta)$ is indeed a probability distribution, as 
\( \int_{\er^d} f_{\cM^*}(x|\psi,\theta)\dx = \int_{\er^d} f_\cM(z|\theta){\rm d}z=1\).
The fact that the (positive) Jacobian $\det(\ddx\psi(x))$ is involved in \refeq{psiM} will be an important specificity of this paper compared to previous work.


\subsection{Registration as an inference problem}

Given an observed point set in $\er^d$, $X=\{x_n\}_{n=1\dots N}$, we can now reformulate our registration problem as a probabilistic inference problem~: \emph{to find GMM parameters $\theta$ and diffeomorphism $\psi$ that maximize the likelihood of having generated the points in $X$}. For this, we make a number of classic assumptions \cite{Chui2000,Granger2002,Myronenko2010}. First, that the points $\{x_n\}_{n=1\dots N}$ have been generated as independent samples from distribution $\cM^*(\psi,\theta)$. Second, that there is no probabilistic prior over GMM parameters $\theta$ (informally, just set $\pP(\theta)=1$). 
Third, this article considers diffeomorphic mappings $\psi$ belonging to an LDDMM group $\Gmap$, detailed below. Every mapping $\psi\in\Gmap$ is naturally associated to a number $\DG(\psi)>0$ quantifying the amount of deformation induced by $\psi$ (\refeq{distpsi} below). Hence, we can set a probabilistic prior over mappings $\psi\in\Gmap$ to be
\begin{equation}
\label{eq:Gprior}
\pP(\psi) \sim \exp(-\lambda \DG(\psi))
\end{equation}
where $\lambda>0$ is a model parameter controlling the amount of deformation allowed for $\psi$, and the normalization is unimportant. Finally, the values of $\psi$ and $\theta$ are assumed to be independent, i.e., $\pP(\psi|\theta)=\pP(\psi)$.
Under all these assumptions, the total likelihood of data and parameters as given by Bayes' law writes
\( \pP(X,\psi,\theta) = \pP(\theta)\pP(\psi|\theta)\pP(X|\psi,\theta)=\pP(\psi) \prod_{n=1}^N \pP(x_n|\psi,\theta) \).
We are thus led to solve the following maximum likelihood optimization problem~:
\begin{equation}
\label{eq:maxlik}
\max_{\substack{\theta\\\psi\in\Gmap}} \;\; \pP(\psi) \prod_{n=1}^N f_{\cM^*}(x_n|\psi,\theta)
\end{equation}
with $f_{\cM^*}(x|\psi,\theta)$ given by \refeq{psiM} and $\pP(\psi)$ given by \refeq{Gprior}.
Note that problem \refno{maxlik} can easily be generalized to multiple point sets~: see \refeq{maxlikCo} below. The main exposition will be for a single point set, only to lighten notations.

\paragraph{EM resolution.}
The optimization problem \refeq{maxlik} is typically solved with an Expectation-maximization (EM) algorithm \cite{Chui2000,Granger2002,Myronenko2010,Evangelidis2017}. This is the method of choice for maximum-likelihood problems involving mixture distributions~; it provides faster and more robust convergence than naive gradient-based methods \cite{Bishop2006}. Following the classic EM procedure, we introduce responsibility variables $\gamma_{nc}>0$ between each data point $n$ and GMM component $c$, constrained by $\forall n,\; \sum_c \gamma_{nc}=1$, and derive from \refeq{maxlik} the following EM free energy~:
\begin{align}
\cF(\gamma,\theta,\psi) &:=  \sum_{n=1}^N\sum_{c=1}^C \gamma_{nc}\left(\frac{|\psi(x_n)-\mu_c|^2}{2\sigma^2}+\log\frac{\sigma^d}{\pi_c}+\log\gamma_{nc}\right)\nonumber\\
& - \sum_{n=1}^N \log\det(\ddx\psi(x_n)) + \lambda\DG(\psi) \label{eq:freeEM}
\end{align}
The EM algorithm proceeds to repeated partial minimizations of $\cF$ with respect to $\gamma$, then $\theta$, then $\psi$, circularly,
until a local minimum is reached. 
At the minimum, the obtained values for $\theta$ (GMM parameters) and $\psi$ (diffeomorphism) also constitute a local maximum of the original likelihood problem, \refeq{maxlik}.

The minimizations of $\cF$ w.r.t. $\gamma$ and $\theta$ are classic computations from the GMM model \cite{Bishop2006,Chui2000,Granger2002,Myronenko2010,Evangelidis2017}. In the \emph{E step}, solving $\partial_\gamma \cF=0$ (subject to $\forall n,\; \sum_c \gamma_{nc}=1$) yields the following update rule for $\gamma$~:
\begin{equation}
\label{eq:gammaUp}
\gamma_{nc} = \frac{\pi_c e^{-|\psi(x_n)-\mu_c|^2/2\sigma^2}}{\sum_{c'} \pi_{c'} e^{-|\psi(x_n)-\mu_{c'}|^2/2\sigma^2}}
\end{equation}
In the \emph{M step}, solving $\partial_\sigma \cF=0$, resp. $\partial_{\mu_c}\cF=0$, resp. $\partial_{\pi_c}\cF=0$ subject to $\sum_c\pi_c=1$,  yields the respective update rules for the GMM parameters (of which, only those pertaining to the problem's free parameters $\theta$ should be applied)~:
\begin{align}
\sigma^2 = \frac1{dN}\sum_{n,c} \gamma_{nc}|\psi(x_n)-\mu_c|^2,\quad
\mu_c = \frac{\sum_n\gamma_{nc}\psi(x_n)}	{\sum_n \gamma_{nc}},\quad
\pi_c = \frac1N \sum_n\gamma_{nc}\label{eq:piUp}
\end{align}

In contrast, the minimization of \refeq{freeEM} w.r.t. $\psi$ yields a novel registration problem in the LDDMM framework (\refeq{Epsi} below), which we present now.


\subsection{LDDMM registration with logdet term}
\label{LDDMMlogdet}

Due to space constraints, we assume some prior acquaintance with the LDDMM framework \cite{Joshi2000,Beg2005,Miller2006}, only listing rapidly its main elements required in the sequel.

\paragraph{Space $\HV$ of vector fields.}
The theory starts by defining a functional space $\HV$ of vector fields over $\er^d$, a Reproducing Kernel Hilbert Space (RKHS) whose inner product will be noted $\scalV{v}{w}$, and associated norm $\normV{v}^2:=\scalV{v}{v}$. 
The reproducing kernel of $\HV$ is assumed to be translation invariant~; we note it as $\kV_z(y):=\kV(y-z)$, with $\kV~:\er^d\to\er$ the chosen, radial, kernel function. 
Evaluation functionals \emph{and their first spatial derivatives} are assumed to be continuous on $\HV$~; hence we have for every $w\in\HV$, indices $i,j\in\oneto{d}$, and $z\in\er^d$,
\begin{equation}
\label{eq:RKHS}
w^i(z)=\scalV{K_z e_i}{w} \quad\textrm{and}\quad \partial_jw^i(z)=\scalV{(\partial_jK_z) e_i}{w} 
\end{equation}
where $w^i$ denotes the $i$-th component of vector field $w$, $e_i$ is the $i$-th elementary vector of $\er^d$, and $\partial_j$ is spatial derivation w.r.t. to the $j$-th component.

\paragraph{LDDMM diffeomorphism group and geodesics.}
We note $\Gmap$ the subgroup of diffeomorphisms on $\er^d$ that can be obtained from the flow of vector fields belonging to $\HV$. Precisely, $\psi\in\Gmap$ iif. $\psi=\phi_1$  where $t\mapsto\phi_t$ is the flow (i.e., $\phi_0={\rm Id}$) associated to the ODE
\begin{equation}
\label{eq:flow}
\partial_t\phi_t = \V_t\circ\phi_t
\end{equation}
with $t\mapsto\V_t$ a time-evolving vector field with values in $\HV$, and sufficient regularity \cite{Miller2006}.
In the sequel, notation $\V_t$ will always represent the time-derivative of $\phi_t$ according to \refeq{flow}, without necessarily reminding it.

$\Gmap$ is then equipped with the metric inherited from $\HV$~: the squared distance along each trajectory $t\mapsto\phi_t$ is defined as $\frac12\int_{t=0}^1 \normV{\V_t}^2\dt$.
This allows to define the concept of an LDDMM \emph{geodesic}, i.e., a trajectory $t\mapsto\phi_t$ minimizing the squared distance, given imposed starting point $\phi_0={\rm Id}$ and endpoint $\phi_1=\psi$. These curves can be characterized with a classic Euler-Lagrange perturbative approach \cite{Beg2005,Miller2006}, yielding the following \emph{geodesic equation} on the trajectory~:
\begin{equation}
\label{eq:geodesic}
\forall w\in\HV,\quad \partial_t\scalV{\V_t}{w}=\scalV{\V_t}{\Lie_{\V_t}w}
\end{equation}
with the Lie derivative $\Lie_vw:=\ddx w.v-\ddx v.w$. 
Concretely, \refeq{flow}-\refno{geodesic} allow to compute the full geodesic trajectory $t\mapsto\phi_t$ from the initial value $\V_0\in\HV$.

Setting $w=\V_t$ in \refeq{geodesic}, we obtain $\partial_t\normV{\V_t}^2=0$, so $\normV{\V_t}$ along a geodesic is constant. This leads to introduce, for every $\psi\in\Gmap$, the functional
\begin{equation}
\label{eq:distpsi}
\DG(\psi):=\frac12\normV{\V_0}^2 \quad\left(=\frac12\normV{\V_t}^2,\quad\forall t\in[0,1]\right)
\end{equation}
for the only geodesic $t\mapsto\phi_t$ such that $\phi_0={\rm Id}$ and $\phi_1=\psi$. 
The value of $\DG(\psi)$ measures the (minimal) squared distance in $\Gmap$ from ${\rm Id}$ to $\psi$. This is the measure that we use as a probabilistic prior on $\psi$~: see \refeq{Gprior} above.



\paragraph{Energy functional.}
Returning to our registration problem, and focusing on the dependency of $\cF$ in \refeq{freeEM} with respect to $\psi$, we have to minimize
\begin{equation}
\cF(\psi) = {\cal C}+\sum_{n=1}^N\frac{|\phi_1(x_n)-y_n|^2}{2\sigma^2}+\int_{t=0}^1\Big(\frac\lambda2\normV{\V_t}^2-\sum_{n=1}^N\div(\V_t)_{\phi_t(x_n)}\Big)\dt \label{eq:Epsi}
\end{equation}
where ${\cal C}$ denotes terms in $\cF$ independent of $\psi$, and we set $y_n:=\sum_c \gamma_{nc}\mu_c$.
Notation $t\mapsto(\phi_t,\V_t)$ refers to the only geodesic such that $\phi_0={\rm Id}$ and $\phi_1=\psi$, and we use the important fact that
\[ \partial_t \log\det(\ddx\phi_t)=\Tr((\ddx\phi_t)^{-1}\partial_t\ddx\phi_t)=\Tr(\ddx\V_t)=\div(\V_t). \]

If not for the divergence term, \refeq{Epsi} would correspond to the classic ``landmark registration'' problem in the LDDMM framework \cite{Joshi2000}. The resolution here will thus be very similar, but with additional terms. Tracing back our equations, the divergence term comes from the Jacobian $\det(\ddx\psi(x))$ in \refeq{psiM}, and thus reflects our modeling of the registration as a \emph{probabilistic inference}. 


\paragraph{Finite-dimensional geodesic ODE.}
We now let $\beta:=\lambda^{-1}$, and $\tji_n(t):=\phi_t(x_n)$ the $N$ trajectories of the data points under the flow. It can be shown that if $\psi$ is a local optimum of $\cF$ in \refeq{Epsi}, there exist $N$ vector-valued weight functions $t\mapsto a_n(t)\in\er^d$, such that the corresponding geodesic's vector field $\V_t$ is of the form
\begin{equation}
\label{eq:Vt}
\V_t(z) = \sum_{n=1}^N \left[a_n(t)\kV(z-\tji_n(t)) - \beta (\nabla\kV)(z-\tji_n(t))\right]
\end{equation}
or equivalently, through the RKHS property \refeq{RKHS},
\begin{equation}
\label{eq:Mt}
\forall w\in\HV,\quad \scalV{\V_t}{w} = \sum_{n=1}^N \left[a_n(t)^\top w_{\tji_n(t)} + \beta \div(w)_{\tji_n(t)}\right]
\end{equation}
A rapid, heuristic explanation is that $\dualV{\V_1}$ has to be of the form \refeq{Mt} when $\psi$ is an optimum of $\cF$ in \refeq{Epsi}, and then the form \refeq{Vt}-\refno{Mt} is preserved under the geodesic equation \refno{geodesic}.
Specifically, injecting \refeq{Vt}-\refno{Mt} into the geodesic equation \refno{geodesic}, and after some simplifications, we find that trajectories $\tji_n(t)$ and vector weights $a_n(t)$ along a geodesic must be satisfy the following Hamiltonian ODE~:
\begin{align}
\forall n,\quad \ddt \tji_n = (\partial_{a_n}H)(\tji,a) 
\quad,\quad 
\ddt a_n &= -(\partial_{\tji_n}H)(\tji,a) \label{eq:angeodesic}
\end{align}
with initial conditions $\tji_n(0)=x_n$ (data points), $a_n(0)=a_n^0$ (for now, a free parameter), and the Hamiltonian function
\begin{equation*}
H(\tji,a) := \frac12 \sum_{n,m}\left[(a_m^\top a_n) K+\beta(a_n-a_m)^\top \nabla K -\beta^{2} (\Delta K)\right](\tji_m-\tji_n) 
\end{equation*}
Equation \refno{angeodesic} provides a concrete embodiment of equations \refno{flow}-\refno{geodesic} into a finite dimensional ODE on $2dN$ scalar variables. 
As in every Hamiltonian system, we recover that the solution verifies $H(\tji(t),a(t))=\frac12\scalV{\V_t}{\V_t}={\rm constant}=\DG(\psi)$, in accordance with general LDDMM principles.

\paragraph{Geodesic shooting.} Through ODE \refno{angeodesic}, the final diffeomorphism $\psi=\phi_1$ becomes a function of the initial momentum variables $a^0=(a_n^0)_{n=1\dots N}$. Precisely, injecting \refeq{Vt}-\refno{Mt} in \refeq{Epsi}, we obtain $\cF(\psi)={\cal C}+E(a^0)$ for the function
\begin{equation}
\label{eq:Ea0}
E(a^0) := \sum_{n}\frac{|\tji_n(1)-y_n|^2}{2\sigma^2}+\frac\lambda2\int_{t=0}^1\sum_{m,n}\left[(a_m^\top a_n)\kV+\beta^{2}\Delta \kV\right](\tji_m-\tji_n)\dt
\end{equation}
where $a(t),\tji(t)$ are implicit functions of $a^0$ through the geodesic equation \refno{angeodesic}. 
The gradient $\nabla_{a^0}E$ can also be 
estimated~: differentiating \refeq{angeodesic} w.r.t. $a^0$ yields a so-called \emph{auxiliary} linear ODE on the quantities $(\partial_{a^0}\tji)(t)$, $(\partial_{a^0}a)(t)$, and the differentiation of \refeq{Ea0} w.r.t. $a^0$ involves precisely these quantities $\partial_{a^0}\tji$, $\partial_{a^0}a$. Furthermore, this computation can be done automatically by numerical libraries such as PyTorch equipped with automatic differentiation.

Hence, a local minimum for $E(a^0)$ can be found with a \emph{geodesic shooting} procedure~: start with an initial guess for $a^0$, numerically implement ODE \refno{angeodesic} to estimate $E(a^0)$, and the auxiliary computations to estimate $\nabla_{a^0}E$. This allows to modify $a^0$ according to some version of gradient descent, and the whole shooting procedure can be repeated, until a local minimum of $E$ is found.
This minimum solves the partial minimization of $\cF$ in \refeq{freeEM} w.r.t. $\psi$.



\section{Numerical applications}

\paragraph{Implementation.}
We coded in Python, using libraries PyTorch and KeOps \cite{KeOps2021}.
Some elements of code (LDDMM implementation, visualizations in Figures 1-2) were adaptated from KeOps tutorials.
RKHS kernel was chosen as $\kV(z)=\exp(-|z|^2/2\tau^2)$ with $\tau=0.2$, and LDDMM regularization constant as $\lambda=500$.
The geodesic ODE \refeq{angeodesic} was numerically integrated with Ralston's method over 10 discrete time steps.
The gradient $\nabla_{a^0}E$ of \refeq{Ea0} was then estimated automatically by 
back-propagating PyTorch's autograd algorithm through the computations, 
and input into PyTorch's L-BFGS algorithm to target a local minimum of $E(a^0)$.
All code is at \href{https://github.com/AdrienWohrer/diff-icp}{https://github.com/AdrienWohrer/diff-icp}.


\paragraph{Warping to a known GMM distribution.}
In a first experiment, a known GMM model $\cM_g$ (Figure 1a) and unknown diffeomorphism $\psi_g$ are used to generate a warped point set $(x_n)_{n=1\dots 100}$ (Figure 1b), and the goal is to estimate the unknown diffeomorphism.
In our notations, $\theta=\{\}$ (no GMM parameters to optimize), and we seek a mapping $\psi$ to maximize the likelihood in \refeq{maxlik}. 
This is a diffeomorphic generalization of classic ``probabilistic ICP'' algorithms for two-set registration, \refeq{twoset}, in which the second point set constitutes the centroids of the GMM model  \cite{Chui2000,Granger2002,Myronenko2010} 
(except that these algorithms also optimize the GMM variance parameter $\sigma$, whereas we keep it fixed in this simple illustration).

This optimal $\psi$ is found by looping repeatedly through the E-step update \refeq{gammaUp}, and the minimization of \refeq{Epsi} w.r.t. $\psi$ (obtained by minimizing \refeq{Ea0} w.r.t. $a^0$).
Before the first loop, variables $a^0$ are initialized to represent an initial mapping $\psi\simeq{\rm Id}$. After a number of loops, convergence is achieved, providing a warping of the point set back to its generative GMM model (Figure 1c,d).



\begin{figure}[p]
\caption{Warping to a known GMM distribution (see text).}
\centerline{\includegraphics[width=1.2\textwidth]{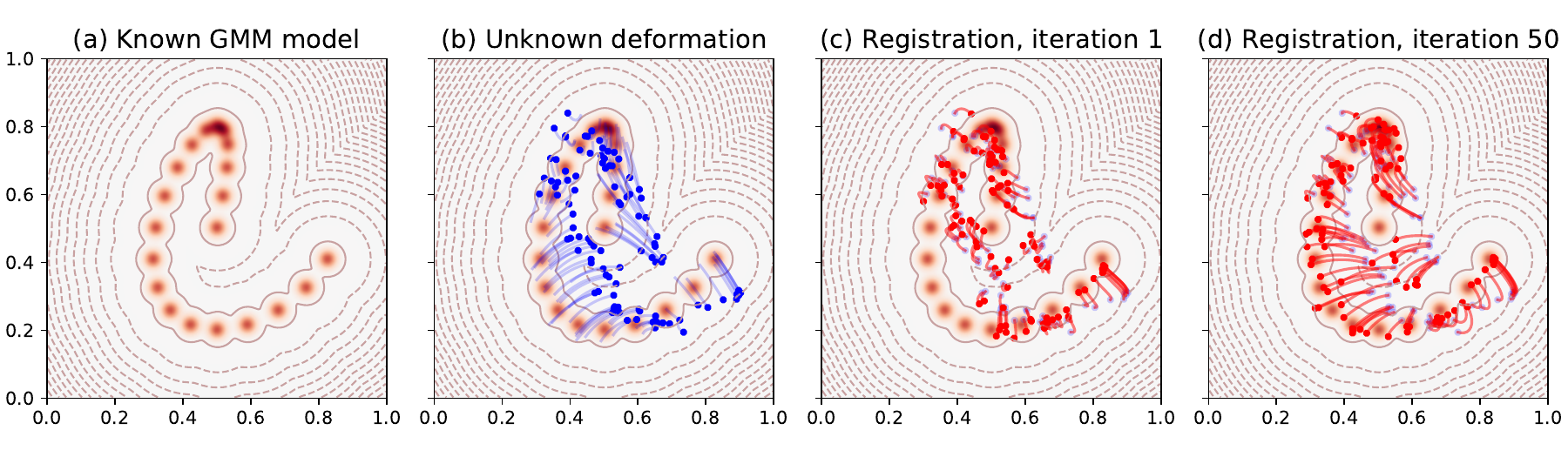}}
\end{figure}
\begin{figure}[p]
\caption{Multiple point set diffeomorphic registration (see text).}
\vspace{1mm}

\centerline{(a) Multiple point set data}
\centerline{\includegraphics[width=1.2\textwidth]{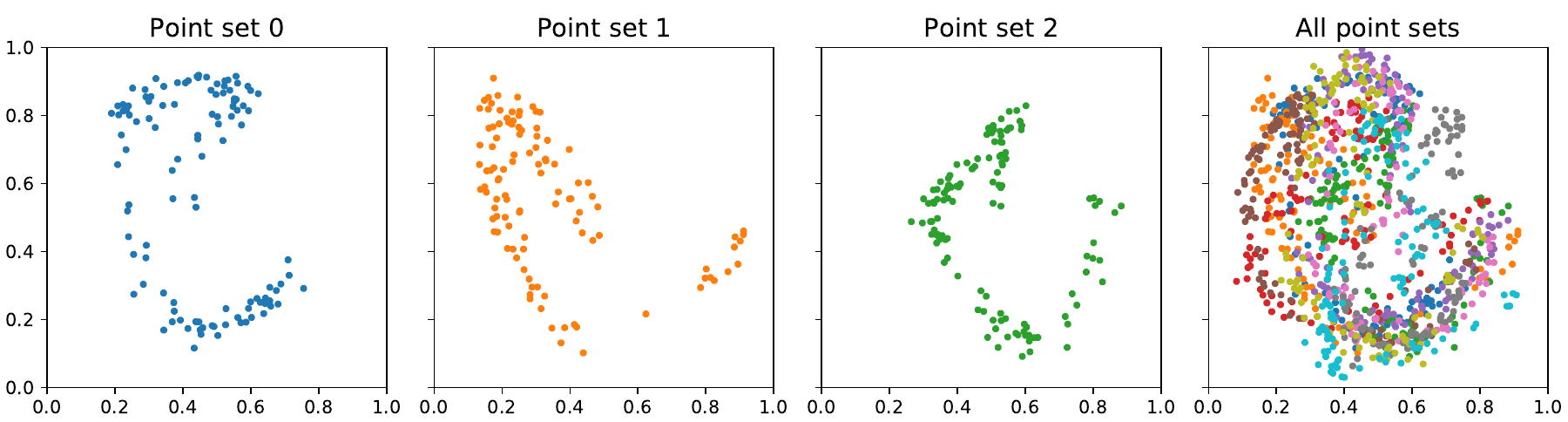}}

\centerline{(b) Diffeomorphic ICP algorithm}
\centerline{\includegraphics[width=1.2\textwidth]{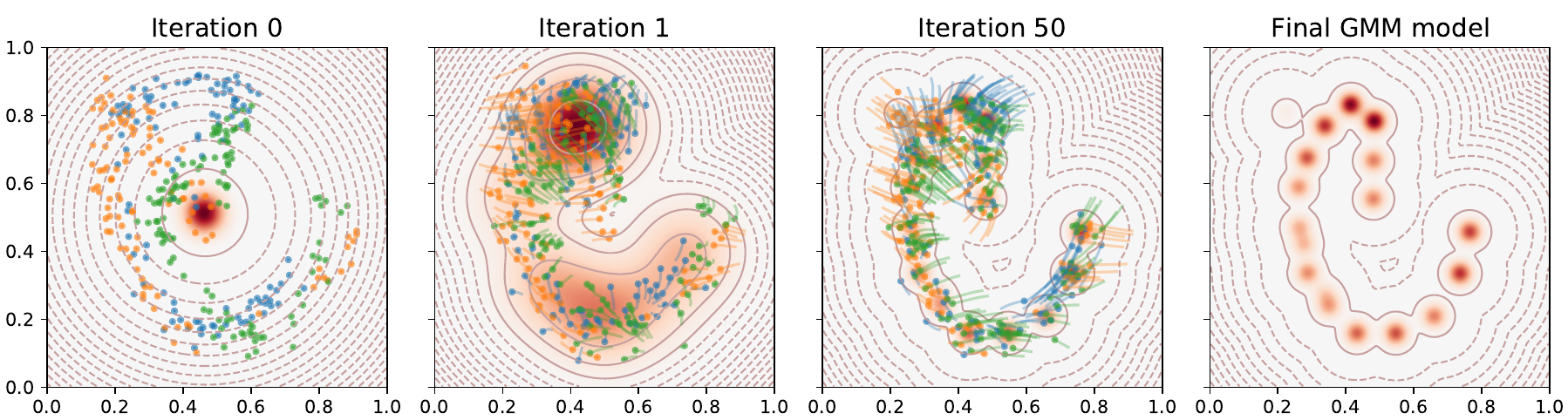}}

\centerline{(c) Simpler algorithm without logdet term}
\centerline{\includegraphics[width=1.2\textwidth]{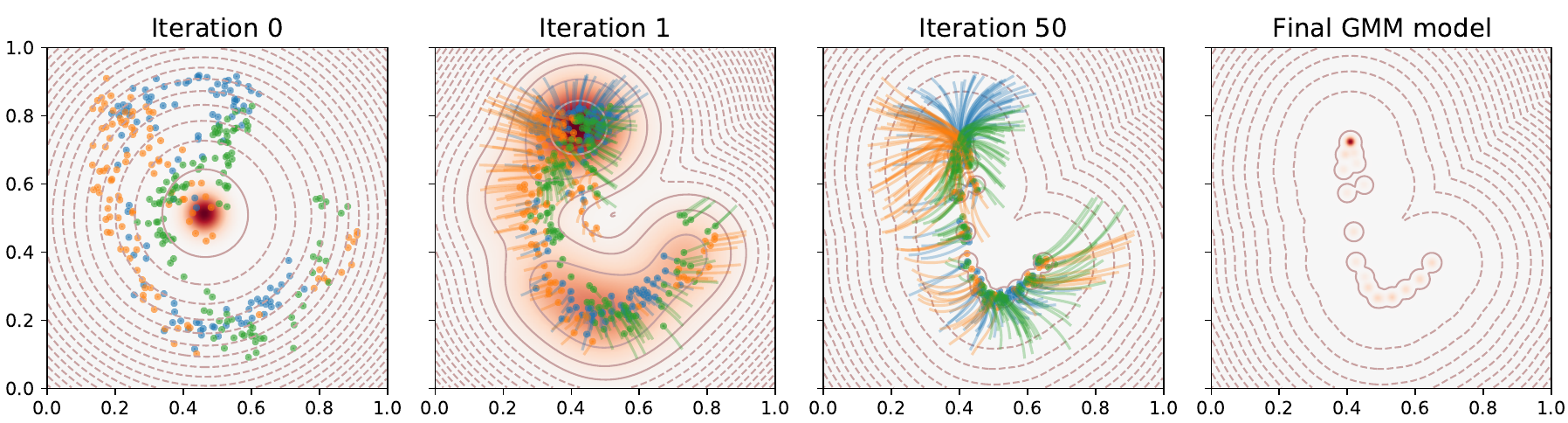}}
\end{figure}

\paragraph{Registration of multiple point sets.}


In a second experiment, we extend the model to perform registration of multiple point sets, \refeq{multiset}. The goal is now to register each point set $X^{(k)}=\{x_{n}^{(k)}\}_{n=1\dots N_k}$ with its dedicated mapping $\psi_k$, to a common space where all data points can be fitted with a single GMM model, that must also be characterized. That is, in our notations,
\begin{equation}
\label{eq:maxlikCo}
\max_{\substack{\theta\\\forall k,\; \psi_k\in\Gmap}} \;\; \prod_{k=1}^K \pP(\psi_k) \left(\prod_{n=1}^{N_k} f_{\cM^*}(x_{n}^{(k)}|\psi_k,\theta)\right)
\end{equation}
 with $\theta=\{\sigma,\mu_c,\pi_c\}$ (all GMM parameters must be optimized).
 This is a diffeomorphic generalization of the rigid registration algorithm for multiple point sets proposed by \cite{Evangelidis2017}.
In a typical application, point set $k$ could represent some anatomical features from patient number $k$, and the GMM $\cM(\theta)$ recovered by the algorithm represents a  \emph{statistical atlas} of these features across patients, as has been proposed, e.g., in the context of image registration \cite{Joshi2004}.

The resolution algorithm now consists in looping through the two following stages until convergence~:
\begin{enumerate}\itemsep2mm
\item Optimize $\cM(\theta)$ to the set of \emph{all} warped points $\psi_k(x^{(k)}_n)$, given the current mappings $\psi_k$. We achieve this by looping 10 times through \refeq{gammaUp}-\refno{piUp}, with  \refeq{piUp} running over all data points, i.e., $\sum_n$ means $\sum_{k=1}^K\sum_{n=1}^{N_k}$.
\item Update each mapping $\psi_k$ to minimize \refeq{Epsi}, given the current GMM $\cM(\theta)$. This can be done independently for each $k$.
\end{enumerate}

We tested this method on $K=10$ point sets generated from the same GMM as Figure 1a, now considered unknown (Figure 2a). After a number of loops, all mappings $\psi_k$ and the joint GMM $\cM(\theta)$ converge to an equilibrium (Figure 2b). In particular, the inferred GMM (``statistical atlas'') correctly recovers the shape of the generating GMM.

We also tested a modified algorithm, replacing the minimization of \refeq{Epsi} by the classic LDDMM ``landmark registration'' algorithm \cite{Joshi2000}, which is recovered by setting $\beta=0$ in \refeq{Vt} and following. In this case, the obtained mappings $\psi_k$ excessively shrink the point sets (Figure 2c), as this allows to artificially minimize the quadratic error in \refeq{Epsi}. This demonstrates the need of the additional divergence term in \refeq{Epsi} to obtain a well-posed algorithm in general.

\section{Conclusion}

We have proposed a generalization of probabilistic ICP algorithms based on the GMM distribution \cite{Chui2000,Granger2002,Myronenko2010}, and their generalization to multiple point sets \cite{Evangelidis2017}, to incorporate diffeomorphic registration in the LDDMM framework. 
First experiments on synthetic data suggest that the algorithm can correctly register two point sets diffeomorphically (Figure 1), or retrieve a common shape underlying multiple point sets (Figure 2). Future work will have to confirm these results more quantitatively, and include comparisons with previous algorithms. 

The algorithm, presented here in its ``raw'' form, could easily accomodate the numerous variations that have been proposed to improve performance of previous ICP algorithms~: introducing an additional mixture component to handle outliers \cite{Chui2000,Evangelidis2017}, replacing the ICP association rule by one-to-one associations (as in the RPM model with \emph{softassign} \cite{Chui2003}), controlling GMM parameter $\sigma$ through an external annealing program \cite{Chui2000,Chui2003,Granger2002}, etc.

We also intend to investigate the roles of the different model meta-parameters~: number of GMM classes $C$, LDDMM regularisation strength $\lambda$, RKHS kernel spatial scape $\tau$, and to provide some heuristics for fixing their values. Finally, we will concretely apply the algorithm to the multiple registration of deep brain structures in a cohort of surgically implanted patients. 





%
%
%
\bibliographystyle{splncs04}

\begin{thebibliography}{10}
\providecommand{\url}[1]{\texttt{#1}}
\providecommand{\urlprefix}{URL }
\providecommand{\doi}[1]{https://doi.org/#1}

\bibitem{Beg2005}
Beg, M.F., Miller, M.I., Trouv\'e, A., Younes, L.: Computing large deformation
  metric mappings via geodesic flows of diffeomorphisms. International journal
  of computer vision  \textbf{61},  139--157 (2005)

\bibitem{Besl1992}
Besl, P.J., McKay, N.D.: Method for registration of 3-d shapes. In: Sensor
  fusion IV: control paradigms and data structures. vol.~1611, pp. 586--606.
  Spie (1992)

\bibitem{Bishop2006}
Bishop, C.M.: Pattern recognition and machine learning. Springer Verlag, New
  York, USA (2006)

\bibitem{KeOps2021}
Charlier, B., Feydy, J., Glaunès, J.A., Collin, F.D., Durif, G.: Kernel
  operations on the gpu, with autodiff, without memory overflows. Journal of
  Machine Learning Research  \textbf{22}(74), ~1--6 (2021)

\bibitem{Chui2000}
Chui, H., Rangarajan, A.: A feature registration framework using mixture
  models. In: Proceedings IEEE Workshop on Mathematical Methods in Biomedical
  Image Analysis. MMBIA-2000 (Cat. No.PR00737). pp. 190--197 (2000)

\bibitem{Chui2003}
Chui, H., Rangarajan, A.: A new point matching algorithm for non-rigid
  registration. Computer Vision and Image Understanding  \textbf{89}(2),
  114--141 (2003)

\bibitem{Evangelidis2017}
Evangelidis, G.D., Horaud, R.: Joint alignment of multiple point sets with
  batch and incremental expectation-maximization. IEEE transactions on pattern
  analysis and machine intelligence  \textbf{40}(6),  1397--1410 (2017)

\bibitem{Granger2002}
Granger, S., Pennec, X.: Multi-scale em-icp: A fast and robust approach for
  surface registration. In: European conference on computer vision. pp.
  418--432. Springer (2002)

\bibitem{Joshi2004}
Joshi, S., Davis, B., Jomier, M., Gerig, G.: Unbiased diffeomorphic atlas
  construction for computational anatomy. NeuroImage  \textbf{23},  S151--S160
  (2004)

\bibitem{Joshi2000}
Joshi, S.C., Miller, M.I.: Landmark matching via large deformation
  diffeomorphisms. IEEE transactions on image processing  \textbf{9}(8),
  1357--1370 (2000)

\bibitem{Maiseli2017}
Maiseli, B., Gu, Y., Gao, H.: Recent developments and trends in point set
  registration methods. Journal of Visual Communication and Image
  Representation  \textbf{46},  95--106 (2017)

\bibitem{Miller2006}
Miller, M.I., Trouvé, A., Younes, L.: Geodesic shooting for computational
  anatomy. Journal of mathematical imaging and vision  \textbf{24},  209--228
  (2006)

\bibitem{Myronenko2010}
Myronenko, A., Song, X.: Point set registration: Coherent point drift. IEEE
  transactions on pattern analysis and machine intelligence  \textbf{32}(12),
  2262--2275 (2010)

\bibitem{Zhu2019}
Zhu, H., Guo, B., Zou, K., Li, Y., Yuen, K.V., Mihaylova, L., Leung, H.: A
  review of point set registration: From pairwise registration to groupwise
  registration. Sensors  \textbf{19}(5), ~1191 (2019)

\end{thebibliography}


%










\end{document}